\documentclass[12pt]{article}
\usepackage{amsfonts}
\usepackage{amssymb}

\def\hybrid{\topmargin -20pt    \oddsidemargin 0pt
        \headheight 0pt \headsep 0pt
        \textwidth 6.25in       
        \textheight 9.5in       
        \marginparwidth .875in
        \parskip 5pt plus 1pt   \jot = 1.5ex}

\hybrid

\def\baselinestretch{1.2}

\catcode`\@=11

\def\marginnote#1{}
\newcount\hour
\newcount\minute
\newtoks\amorpm
\hour=\time\divide\hour by60
\minute=\time{\multiply\hour by60 \global\advance\minute by-\hour}
\edef\standardtime{{\ifnum\hour<12 \global\amorpm={am}%
        \else\global\amorpm={pm}\advance\hour by-12 \fi
        \ifnum\hour=0 \hour=12 \fi
        \number\hour:\ifnum\minute<10 0\fi\number\minute\the\amorpm}}
\edef\militarytime{\number\hour:\ifnum\minute<10 0\fi\number\minute}

\def\draftlabel#1{{\@bsphack\if@filesw {\let\thepage\relax
   \xdef\@gtempa{\write\@auxout{\string
      \newlabel{#1}{{\@currentlabel}{\thepage}}}}}\@gtempa
   \if@nobreak \ifvmode\nobreak\fi\fi\fi\@esphack}
        \gdef\@eqnlabel{#1}}
\def\@eqnlabel{}
\def\@vacuum{}
\def\draftmarginnote#1{\marginpar{\raggedright\scriptsize\tt#1}}

\def\draft{\oddsidemargin -.5truein
        \def\@oddfoot{\sl preliminary draft \hfil
        \rm\thepage\hfil\sl\today\quad\militarytime}
        \let\@evenfoot\@oddfoot \overfullrule 3pt
        \let\label=\draftlabel
        \let\marginnote=\draftmarginnote
   \def\@eqnnum{(\theequation)\rlap{\kern\marginparsep\tt\@eqnlabel}%
\global\let\@eqnlabel\@vacuum}  }

\def\preprint{\twocolumn\sloppy\flushbottom\parindent 2em
        \leftmargini 2em\leftmarginv .5em\leftmarginvi .5em
        \oddsidemargin -.5in    \evensidemargin -.5in
        \columnsep .4in \footheight 0pt
        \textwidth 10.in        \topmargin  -.4in
        \headheight 12pt \topskip .4in
        \textheight 6.9in \footskip 0pt
        \def\@oddhead{\thepage\hfil\addtocounter{page}{1}\thepage}
        \let\@evenhead\@oddhead \def\@oddfoot{} \def\@evenfoot{} }

\def\numberbysection{\@addtoreset{equation}{section}
        \def\theequation{\thesection.\arabic{equation}}}

\def\underline#1{\relax\ifmmode\@@underline#1\else
        $\@@underline{\hbox{#1}}$\relax\fi}

\def\titlepage{\@restonecolfalse\if@twocolumn\@restonecoltrue\onecolumn
     \else \newpage \fi \thispagestyle{empty}\c@page\z@
        \def\thefootnote{\fnsymbol{footnote}} }

\def\endtitlepage{\if@restonecol\twocolumn \else \newpage \fi
        \def\thefootnote{\arabic{footnote}}
        \setcounter{footnote}{0}}  

\catcode`@=12
\relax

\def\figcap{\section*{Figure Captions\markboth
        {FIGURECAPTIONS}{FIGURECAPTIONS}}\list
        {Figure \arabic{enumi}:\hfill}{\settowidth\labelwidth{Figure
999:}
        \leftmargin\labelwidth
        \advance\leftmargin\labelsep\usecounter{enumi}}}
 \relax
\def\tablecap{\section*{Table Captions\markboth
        {TABLECAPTIONS}{TABLECAPTIONS}}\list
        {Table \arabic{enumi}:\hfill}{\settowidth\labelwidth{Table
999:}
        \leftmargin\labelwidth
        \advance\leftmargin\labelsep\usecounter{enumi}}}
 \relax
\def\reflist{\section*{References\markboth
        {REFLIST}{REFLIST}}\list
        {[\arabic{enumi}]\hfill}{\settowidth\labelwidth{[999]}
        \leftmargin\labelwidth
        \advance\leftmargin\labelsep\usecounter{enumi}}}
 \relax
\makeatletter
\newcounter{pubctr}
\def\publist{\@ifnextchar[{\@publist}{\@@publist}}
\def\@publist[#1]{\list
        {[\arabic{pubctr}]\hfill}{\settowidth\labelwidth{[999]}
        \leftmargin\labelwidth
        \advance\leftmargin\labelsep
        \@nmbrlisttrue\def\@listctr{pubctr}
        \setcounter{pubctr}{#1}\addtocounter{pubctr}{-1}}}
\def\@@publist{\list
        {[\arabic{pubctr}]\hfill}{\settowidth\labelwidth{[999]}
        \leftmargin\labelwidth
        \advance\leftmargin\labelsep
        \@nmbrlisttrue\def\@listctr{pubctr}}}
 \relax
\makeatother
\newskip\humongous \humongous=0pt plus 1000pt minus 1000pt

\newif\ifdtup

\relax

\def\be{\begin{equation}}
\def\ee{\end{equation}}
\def\ba{\begin{eqnarray}}
\def\ea{\end{eqnarray}}

\def\del{\partial}

\def\r{\rho}
\def\a{\alpha}

\def\b{\beta}

\def\g{\gamma}
\def\G{\Gamma}
\def\d{\delta}
\def\D{\Delta}

\def\m{\mu}
\def\n{\nu}

\def\l{\lambda}

\def\cN{{\cal N}}

\def\cL{{\cal L}}
\def\cR{{\cal R}}

\def\no{\noindent}

\def\qq{\qquad}

\def\IR{\relax{\rm I\kern-.18em R}}

\def \ha {{1\over 2}}

\def \ov {\over}

\def\IR{\relax{\rm I\kern-.18em R}}
\def\inv{^{\raise.15ex\hbox{${\scriptscriptstyle -}$}\kern-.05em 1}}

\def\cL{{\cal L}}
\def\cR{{\cal R}}

\begin{document}

\renewcommand{\theequation}{\arabic{equation}}

\newcommand{\beq}{\begin{equation}}
\newcommand{\eeq}[1]{\label{#1}\end{equation}}
\newcommand{\ber}{\begin{eqnarray}}
\newcommand{\eer}[1]{\label{#1}\end{eqnarray}}
\newcommand{\eqn}[1]{(\ref{#1})}
\begin{titlepage}
\begin{center}

\hfill CERN-TH/2000-261\\
\vskip -.1 cm
\hfill NEIP-00-016\\
\vskip -.1 cm
\hfill hep--th/0010048\\

\vskip .5in

{\large \bf Current correlators in the Coulomb branch of ${\cal N}=4$ SYM}

\vskip 0.4in

{\bf Andreas Brandhuber}${}^{1,}$\footnote{
Since $30^\mathrm{th}$ September 2000:
Department of Physics, CalTech, Pasadena, CA 91125}
and\phantom{x}
{\bf Konstadinos Sfetsos}${}^2$
\vskip 0.1in
{\em ${}^1\!$Theory Division, CERN\\
     CH-1211 Geneva 23, Switzerland\\
\footnotesize{\tt brandhu@mail.cern.ch}}\\
\vskip .2in
{\em ${}^2\!$Institut de Physique, Universit\'e de Neuch\^atel\\
Breguet 1, CH-2000 Neuch\^atel, Switzerland\\
\footnotesize{\tt sfetsos@mail.cern.ch}}\\

\end{center}

\vskip .3in

\centerline{\bf Abstract}

\noindent
We study correlators of ${\cal R}$-symmetry currents in the Coulomb
branch of ${\cal N} = 4$ supersymmetric gauge theory in the large-$N$ limit,
using the AdS/CFT correspondence. In particular, we consider gauge fields
in the presence of gravity and scalar fields parameterizing the coset
$SL(6,\IR)/SO(6)$ in the context of five-dimensional gauged supergravity.
From a ten-dimensional point of view these backgrounds
correspond to continuous D3-brane distributions.
We find the surprising result that all 2-point functions of gauge
currents fall into the same universality class, irrespectively of whether
they correspond to broken or unbroken symmetries. We show that the problem
of finding the spectrum can be mapped into an equivalent Schr\"odinger
problem for supersymmetric quantum mechanics.
The corresponding potential is the supersymmetric partner of the
potential arising in studies of the spectrum for massless scalars and
transverse graviton fluctuations in these backgrounds
and the associated spectra are also identical.
We discuss in detail two examples where these computations can be done
explicitly as in the conformal case.

\vskip .4in
\noindent
CERN-TH/2000-261\\
August 2000\\
\end{titlepage}
\vfill
\eject

\def\baselinestretch{1.2}
\baselineskip 16 pt
\noindent

\section{Introduction}

For several years the dynamics of branes in string theory have been
a fruitful playground to test strong coupling physics of gauge theories.
For instance, the AdS/CFT correspondence \cite{malda,gkp,witten} provides
us with precise prescriptions to calculate
correlation functions, spectra of gauge invariant operators,
Wilson loops and $c$-functions in ${\cal N}=4$ supersymmetric
Yang--Mills (SYM) theory in
four dimensions at large $N$ and large 't Hooft coupling.
The data obtained this way from supergravity can sometimes be
compared with field theory or provide non-trivial predictions
for strongly coupled field theories.
This correspondence can be extended also to theories with
spontaneously or manifestly broken superconformal symmetry.
Such theories arise either by giving vacuum expectation
values to fields \cite{malda}, \cite{kraus}-\cite{bbs}
or by deforming the conformal
theory with relevant operators \cite{gppz1}-\cite{Evans}.
Many of these deformations can be treated efficiently in the context of
five-dimensional gauged supergravity \cite{PPN,GRW}
and the resulting backgrounds have
four-dimensional Poincar\'e invariance and approach $AdS_5$
in the ultraviolet (in a field theory terminology).
Typically, towards the infrared, singularities appear which are
not fully understood and seem to require a proper inclusion of the
string theory dynamics or the use of other methods developed in gravity.

In this letter we study correlation functions of ${\cal R}$-symmetry
currents using the holographic description of large-$N$ gauge theories.
For the conformal case correlation functions for operators
in various representations of the ${\cal R}$-symmetry group $SU(4)\simeq SO(6)$
have been worked out in great detail (see, for instance, \cite{fmmr,dhoker}).
Less is known about
correlators in deformed gauge theories which are described by
more general domain wall solutions of gauged supergravity.
So far mainly scalars have been studied, namely the minimally coupled scalar
\cite{fgpw2,brand1,Anselmi} (which
has the same equation as the transverse traceless graviton modes
\cite{brand2}),
active and inert scalars which parameterize deformations of the $S^5$
\cite{DeWolfe,AFT,massimo}, but also fermionic and abelian vector
field fluctuations for the ${\cal N}=1$ flow of
\cite{gppz2} and the ${\cal N}=4$ Coulomb branch background of
\cite{fgpw2,brand1} have been considered recently in \cite{massimo}.

We will show that for a specific class of
examples this analysis can be extended
to include fluctuations of non-abelian gauge fields which are dual to
${\cal R}$-symmetry currents of the gauge theory.
We make a general connection between the fluctuation equation
and supersymmetric quantum mechanics
and find that, the relevant Schr\"odinger potential, associated with
the spectrum, is
just the supersymmetric partner of the potential arising from
the corresponding massless scalar and transverse graviton-fluctuations
equations. We show also that the corresponding spectra are identical.
It seems plausible to us that this can be extended to the full
set of fields in the supergravity multiplet.
Using the AdS/CFT correspondence we calculate two-point
functions of the symmetry currents in ${\cal N} = 4$ SYM
on the Coulomb branch in two particular cases.\footnote{Other studies
of the Coulomb branch of the ${\cal N} =4$ SYM theory using
the AdS/CFT correspondence can be found in \cite{OthersCoulomb}.}
As expected, we find deviations
from the conformal $1/r^6$ fall-off for large separations $r$.
From the non-analytic part of the correlator in momentum space
we get contributions that are suppressed exponentially for
large separation.

The choice of a particular state on the Coulomb branch breaks the
${\cal R}$-symmetry to a subgroup and therefore one might expect
that broken and unbroken currents behave differently and
in particular one would
expect Goldstone bosons corresponding to the broken symmetry.
From the dual supergravity point of view this symmetry is a
local gauge symmetry
and the massless bosons simply get eaten by the gauge fields and
make them massive via the Higgs mechanism.
Although the equations for broken and unbroken currents look quite
different --- they correspond, respectively,
to massless gauge fields in a curved background
and massive gauge fields --- the associated spectra are identical.
This result is not too surprising since on the Coulomb branch
only conformal symmetry is broken but the currents still reside
in the same supersymmetry multiplet.
However, a small puzzle remains since
the correlator has also an analytic piece
that depends on which of the broken or unbroken currents are
considered. For the two-point function of scalars such analytic
terms give rise to contact terms and are usually dropped, but
in the case of gauge field correlators they give rise to terms of the
form $x_\m x_\n/r^6$, which might be interpreted in field theory as
arising from  Goldstone bosons.
However, we do not find a one to one relation between
broken currents and the presence of these terms in the correlators.
We believe that these analytic terms are unphysical, since the corresponding
mode is non-normalizable, and should be dropped.

The organization of this paper is as follows:
In section 2 we present some background material on gauged supergravity
and calculation of correlators in AdS/CFT.
We also make a general connection between the fluctuation equation
and supersymmetric quantum mechanics.
In section 3 we focus on our two main examples
where calculations can be performed explicitly. We obtain
the exact fluctuation spectrum of gauge fields, and the two-point functions
in momentum and position space.
In section 4 we give a summary of our results
and give some final remarks.

\section{Generalities}

Our starting point is a specific truncation of the ${\cal N} = 8$ gauged
supergravity action \cite{PPN,GRW}
including $SO(6)$ gauge fields $A_{\widehat \m}^{ij}$,
antisymmetric in $i,j$, with
field strength $F_{\widehat \m \widehat \n}^{ij}$,
where $\widehat \m, \widehat \n = 1,2,3,4,z$;
unhatted indices $\mu, \nu =1,2,3,4$ will be used later to
denote Euclidean directions along the boundary at $z=0$.
For notational
convenience we will occasionally
use the collective index $a=1,2,\dots, 15$ to
denote the adjoint representation of $SO(6)$,
instead of $i$ and $j$ or we will omit such an index all together.
Furthermore, scalars in the ${\bf 20^\prime}$ are represented by a
symmetric traceless matrix $M^{ij}$. The action of the supergravity
truncated to these fields
has been constructed in \cite{cvetic} and we follow closely their
conventions.

The Lagrangian density for the relevant fields of five-dimensional
gauged supergravity is
\be
\cL = \cL_{\rm scalar} + \cL_{\rm gauge}\ ,
\label{sugralagr}
\ee
where $\cL_{\rm scalar}$ refers to the pure gravity-scalar sector and
$\cL_{\rm gauge}$ contains the gauge fields and their interaction with the
scalars and gravity.
We first recall some results for the pure gravity-scalar sector
since we are interested to study fluctuations of the gauge fields
in the background of specific solutions of the gravity-scalar sector.
The explicit form of the Lagrangian is
\be
\frac{1}{\sqrt{g}} \cL_{\rm scalar} = {1\ov 4} {\cal R} -
\frac{1}{16} {\rm Tr}
\left( \partial_{\widehat \m} M M^{-1} \partial^{\widehat \m}
M M^{-1} \right) - P
\ ,
\label{actionsc}
\ee
where the potential is
\be
\label{potential}
P =  -{g^2\ov 32} \left[({\rm Tr} M)^2 - 2 {\rm Tr}(M^2)\right] \  ,
\ee
with $g$ being a mass scale.
Alternatively we may use the length scale $R$ via
the relation $g=2/R$.

Supersymmetric
solutions of \eqn{actionsc} preserving 16 supercharges and Poincar\'e symmetry
in four-dimensions have been studied
extensively and they correspond to states on the Coulomb branch of $\cN=4$
SYM theory.
Their interpretation in ten dimensions is simply in terms of a
continuous distribution of D3-branes. For these backgrounds
the matrix of scalar fields can be brought to a diagonal form using a
gauge transformation.
Thus we are left with six scalar fields that parameterize
\be\label{M}
M = {\rm diag} (e^{2 \beta_1}, \ldots, e^{2 \beta_6} )\ ,
\ee
obeying the constraint $\sum_{i=1}^6 \beta_i = 0$.
There are five independent scalar fields, denoted by $\a_I$, $I=1,2,\dots 5$,
and the relation to the $\b_i$'s is given by
$\b_i= \sum^5_{I=1} \l_{iI} \a_I$,
where $\l_{iI}$ is a $6\times 5$ matrix, with rows corresponding to the
fundamental representation of $SL(6,\IR)$; the normalization
conventions can be found in eq. (2.4) of \cite{bbs}.
The metric ansatz reads
\be
ds^2 = e^{2 A(z)} (dz^2 + \eta_{\m\n} dx^\m dx^\n)
= dr^2 + e^{2A(r)} \eta_{\m\n} dx^\m dx^\n\ ,
\label{metriki}
\ee
where the relation between the coordinates $z$ and $r$ is such that
$dr=-e^A dz$. In addition, all scalar fields depend on the variable $r$ or
equivalently $z$.
The most general solution preserving 16 supercharges has been found in
\cite{bakas1} and is conveniently presented in terms of an auxiliary function
$F(g^2 z)$. Specifically, the conformal factor is given by
\be
e^{2 A} =  g^2 (-F^\prime)^{2/3}\ ,
\label{pro1}
\ee
where the prime denotes the derivative with respect to the argument of
$F(g^2z)$. In addition, the profiles of the scalar fields are
\be
e^{2\b_i} = {f^{1/6}\ov F-b_i}\ ,\qq f = \prod_{i=1}^6 (F-b_i)
\ , \qq i=1,2,\dots, 6 \ .
\label{proo}
\ee
The constants of integration are ordered as $b_1\geq b_2 \geq \dots \geq
b_6$ and the function $F$ is constrained to obey the differential equation
\be
(F^\prime)^4 = f\ .
\label{hd3}
\ee
Equating $n$ of the integration constants $b_i$ (or equivalently the
associated scalar fields $\b_i$) corresponds to preserving an $SO(n)$
subgroup of the original $SO(6)$ $\cR$-symmetry group.
We note in passing, that there is a deep connection between solutions
of the gravity-scalar sector of the
five-dimensional gauged supergravity that we
just reviewed, and the theory of algebraic curves and associated Riemann
surfaces to which the differential equation \eqn{hd3} is related
\cite{bakas1,bbs}.

Let us now turn to the part of the Lagrangian containing the gauge fields.
First, we have to replace the partial derivatives in \eqn{actionsc} by
gauge-covariant ones
$\partial_{\widehat \m} M^{ij} \to
\partial_{\widehat \m} M^{ij} + g
(A_{\widehat \m}^{ik} M^{kj} +  A_{\widehat \m}^{jk} M^{ik})$, and,
second, we add the gauge kinetic term
\be\label{gaugeaction}
\frac{1}{\sqrt{g}} \cL_{\rm gauge} =
-\frac{1}{8}
(M^{-1})^{ij}  (M^{-1})^{kl} F^{ik}_{\widehat \m \widehat \n}
F^{jl \widehat \m \widehat \n } \ .
\ee
Since we are interested in two-point functions we only need to keep terms
in \eqn{sugralagr} and \eqn{gaugeaction} which are
quadratic in the gauge fields and the scalar fluctuations in the symmetric
unimodular matrix $M$. Note that although for our solution the matrix $M$ is
diagonal as in \eqn{M}, we have to consider fluctuations along the diagonal
as well as off-diagonal ones.
Using the fact that $M$ is diagonal
\eqn{M} for our backgrounds, we collect all terms that can give quadratic
terms in the fluctuations of the scalars and the gauge fields
\ba
\frac{1}{\sqrt{g}}
\cL_{\rm quad.} & = & - {1\ov 8} e^{-2 (\b_i+\b_j)}
F^{ij}_{\widehat \m \widehat \n} F^{\widehat \m \widehat \n}_{ij}
- {g^2\ov 4} \sinh^2(\b_i -\b_j) A^{ij}_{\widehat \m}
A^{\widehat \m}_{ij}
\nonumber\\
&& -{g\ov 8} {\rm Tr}\Big((\del_{\widehat \m} M M\inv - M\inv
\del_{\widehat \m} M) A^{\widehat \m}\Big)\Big|_{\rm quad.}
\label{actiongor}\\
&& -{1\ov 16} {\rm Tr}(\del_{\widehat \m} M M\inv \del^{\widehat \m} M M\inv)
- P\big|_{\rm quad.} \ .
\nonumber
\ea
The first line above is already quadratic in the gauge field fluctuations.
We emphasize that $F_{\widehat \m \widehat \n}^{ij}=
\del_{\widehat \m} A^{ij}_{\widehat \n} -
\del_{\widehat \n} A^{ij}_{\widehat \m} $ is, for our purposes,
the relevant part of the gauge field strength.
The second line in the above expression is already linear in the
gauge field fluctuaction. Hence, we are supposed to expand it
to linear order in the scalar fluctuations.
Finally, the third line has to be expanded to quadratic order
in the scalar field fluctuations.
In this paper we are only interested in the gauge field
fluctuations which, however, couple to fluctuations of the scalars.
Therefore, it is not a priori correct to simply keep the terms in the
first line in \eqn{actiongor} and drop the rest.
Nevertheless, we will now explain that this procedure gives the
correct result since there is a field redefinition that effectively
decouples the gauge field fluctuations from
those of the scalars.\footnote{We thank M. Bianchi
for prompting us to explain in detail how the decoupling between scalar and
gauge field fluctuations actually works
as well as for other related comments.}
To see that let us expand the second line in \eqn{actiongor} and keep
the linear term in the scalar field fluctuations.
We find that
\ba
&& - {g\ov 8} {\rm Tr}
\Big((\del_{\widehat \m} M M\inv - M\inv \del_{\widehat \m} M)
A^{\widehat \m}\Big) \Big|_{\rm quad.} =
\nonumber\\
&& = {g\ov 8} \Big( (e^{-2\b_j} - e^{-2 \b_i}) \del_{\widehat \m} \d M_{ij}
+2 (e^{-2\b_i}\del_{\widehat \m} \b_j  - e^{-2 \b_j} \del_{\widehat \m}\b_i)
 \d M_{ij} \Big)A^{\widehat \m}_{ij}\ .
\label{ghe}
\ea
From this we immediately deduce that the diagonal fluctuations
$\d M_{ii}$ do not couple to the gauge fields.
A less trivial fact is that the scalar fluctuations in
$\d M_{ij}$ that belong to any unbroken subgroup of $SO(6)$
do not couple to the gauge fields as well. The reason is that in this case
$\b_i=\b_j$, since then the corresponding integration constants in \eqn{proo}
are equal, i.e. $b_i=b_j$.
Hence, let us consider the remaining cases with $\b_i \neq \b_j$ which
arise when the indices $i, j$ belong to the coset.
If we make the field redefinition
\be
A_{\widehat \m}^{ij} \to A_{\widehat \m}^{ij} + {1\ov g}\ \del_{\widehat \m}
\left(\d M_{ij}\ov e^{2\b_i} - e^{2\b_j}\right)\ ,\qq \b_i\neq \b_j\ ,
\label{frfe}
\ee
the mixed terms between scalar and gauge field fluctuations in
\eqn{actiongor} (with the substitution \eqn{ghe} understood)
disappear and the fluctuations decouple.
Note that the field redefinition
\eqn{frfe} acts as an abelian gauge transformation and as such it
leaves the gauge field strength $F_{\widehat \m\widehat \n}^{ij}$
invariant (to the quadratic order we are working).
We emphasize that the field redefinition
\eqn{frfe} does not guarantee that there will be no mixing between
scalar and gauge field fluctuations at the cubic or at some higher order
in the fluctuating fields, but only that the quadratic
fluctuations decouple.
We also note that a similar decoupling mechanism for vector and
scalar fluctuations was found to be at work for the flow of \cite{gppz2}
in \cite{massimo}. There, it was observed that decoupling was achieved
since the gauge field and a (charged) scalar appeared in a
gauge invariant combination.

The field redefinition \eqn{frfe} removes the scalar fluctuations
of $\d M_{ij}$ since it removes terms quadratic in first derivatives
of $\d M_{ij}$ from the Lagrangian. The remaining terms are at most
linear in first derivatives
and of the form $B_{ij} \d M_{ij} \d M_{ij} + B^{\widehat\m}_{ij}
\d M_{ij} \del_{\widehat\m} \d M_{ij} $ for some space-depended $B_{ij}$
and $ B^{\widehat\m}_{ij}$ which are symmetric in $i,j$.
Clearly the derivative-term can be removed
by adding an appropriate total derivative
so that we are left with a non-dynamical field $\d M_{ij}$ corresponding
to no physical degrees of freedom.
What we have is nothing but a manifestation of the
Higgs effect in a curved background. As in flat space-time,
the Goldstone bosons corresponding to the broken gauge symmetries
are eaten by the gauge bosons which then become massive.

Since we are only interested in the gauge field fluctuations we
ignore the scalar fluctuations for the rest of the paper
and concentrate on those for the
gauge fields which, after the redefinition \eqn{frfe}, are described by
the first line in \eqn{actiongor}
\be
\frac{1}{\sqrt{g}}
\cL(A)_{\rm quad.} =- {1\ov 8} e^{-2 (\b_i+\b_j)}
F^{ij}_{\widehat \m \widehat \n} F^{\widehat \m \widehat \n}_{ij}
- {g^2\ov 4} \sinh^2(\b_i -\b_j) A^{ij}_{\widehat \m}
A^{\widehat \m}_{ij}\ .
\label{actiong}
\ee
The second term corresponds to mass terms for the gauge fields,
if the scalar fields $\b_i$ are not equal. This implies that for general
states on the Coulomb branch the bulk gauge symmetry $SO(6)$ is
spontaneously
broken and, hence, that the ${\cal R}$-symmetry group of the field
theory on the boundary is reduced accordingly.
Notice also that the kinetic term for the gauge fields is not canonically
normalized as it gets ``dressed'' by the scalar fields. This will have
important consequences, as we will see.

The equation of motion following from this quadratic action \eqn{actiong}
is:
\be
\d A^{ij}_{\widehat \m} :\qq D_{\widehat \m}
(e^{-2(\b_i+\b_j)} F^{\widehat \m \widehat \n}_{ij})
- g^2 \sinh^2 (\b_i - \b_j) A^{\widehat \n}_{ij}=0\ .
\label{tade}
\ee
In solving these equations we have to distinguish two cases:
First, for the unbroken symmetry (currents),
for which $\b_i=\b_j$, we can use
the gauge symmetry to choose the gauge $A^{ij}_z=0$. This still allows for
restricted gauge transformations with parameters that depend only on the
$x^\m$'s, but not on $z$.
Then, the $\widehat \m=z$ component of the eqs. \eqn{tade} yields the constraint
$\del_z \del_\m A^\m=0$ which allows to eliminate unphysical longitudinal modes
via a restricted gauge transformation.
The equation of motion for the remaining physical (transverse)
modes $A_\m^\bot$ which obey $\partial^\m A_\m^\bot = 0$ is the same
for all components
and can be written as an equation for a scalar field,
which we denote by $\Phi$:
\be
\label{unbroken}
\del_z (e^B \del_z \Phi) + m^2 e^B  \Phi= 0 \ ,
\ee
with the definition
\be
B = A-2 (\b_i +\b_j)\ .
\label{h93}
\ee
To arrive at this equation we have performed a Fourier transform
in the $x^\m$-directions with $k_\m k^\m = -m^2$.\footnote{For
notational simplicity we did not include indices $i,j$ in
defining $B$ in \eqn{h93}. Nevertheless it should be kept in mind that
different choices for the scalar fields $\b_i$ and $\b_j$ lead to
different values for $B$.}

For the broken symmetry currents for which $\b_i\neq\b_j$
we cannot use a gauge symmetry to eliminate degrees of freedom.
In order to calculate the two-point functions we couple the
gauge field to an external source by adding
$-\frac{1}{2} A_{\widehat \mu}^{ij} J^{\widehat \mu}_{ij}$
to the gauge field action \eqn{actiong}. The source is required
to be covariantly conserved, i.e., $D^{\widehat \mu} J_{\widehat \mu}^{ij}=0$.
We choose to decompose the gauge field into transverse modes
$A_\m^\bot$, longitudinal modes $\partial_\mu \xi = A_\m - A_\m^\bot$,
and the component $A_z$.
The equations of motion \eqn{tade} give
\be
\del_z\left(e^B(\del_z A^\bot_\m -\del_\m A_z + \del_z \del_\m \xi)\right)\
+ \ e^B \square A^\bot_\m \  - \
e^C(\square A^\bot_\m +\del_\m \xi) \ = \ e^{3 A} J_z
\label{gh1}
\ee
and
\be
e^B(\square A_z - \del_z\square \xi)  - e^CA_z = e^{3 A} J_z\ ,
\label{gh2}
\ee
where $\square = \eta^{\m\n} \del_\m \del_\n$.
The above coupled system of equations can be further simplified.
By taking the derivatives $\del_\m$ and $\del_z$ in \eqn{gh1}
and \eqn{gh2} respectively, adding up the resulting expressions
and then using the
condition $D^{\widehat \m} J_{\widehat \m}=0$,
we obtain a relation that determines $\xi$ in terms of the component $A_z$,
namely
\be
e^C \square \xi + \partial_z \left(
e^C A_z \right)=0 \ ,
\label{gh3}
\ee
where
\be e^C = g^2 e^{3 A} \sinh^2 (\b_i - \b_j) =  {1\ov 4} g^2 (b_i-b_j)^2
e^{-B}\ .
\label{hgde}
\ee
The first equality defines $C$, whereas the second one follows with
the help of \eqn{proo} and relates $C$ to $B$ which was defined in \eqn{h93}.
Using \eqn{gh3} to solve for $\square \xi$ and then substituting
back the result into \eqn{gh2} we find the equation for the mode $A_z$,
which decouples from the transverse modes:
\be
\label{Azeom}
e^B \square A_z + e^B \partial_z \left( e^{-C}
\partial_z \left( e^C A_z \right) \right) - e^C A_z = e^{3 A} J_z \ .
\ee
With further manipulations using \eqn{Azeom}, we may cast \eqn{gh1} into an
equation for the transverse modes
\be
\label{Atrans}
e^B \square A_\m^\bot + \partial_z \left( e^B
\partial_z A_\m^\bot \right) - e^C A_\m^\bot = e^{3 A} J_\m^\bot \ ,
\label{hgjh2}
\ee
where we have defined the transverse current-source as $J^\bot_\m = (\d_{\m\n}
-\del_\m \del_\n/\square) J_\n$.
In order to compute the two point functions in momentum space
we need solutions of the homogeneous equations \eqn{Azeom} and \eqn{Atrans}.
Actually, after
a Fourier transform in the $x^\m$ brane-directions, we can
write both equations as an equation for a scalar field
\be
\label{broken}
\del_z (e^B \del_z \Phi) +\left( m^2 e^B
-{1\ov 4} g^2 (b_i-b_j)^2 e^{-B} \right) \Phi= 0\ ,
\label{fiiin}
\ee
where we have dropped the source term. Its effect will be
implemented by imposing appropriate boundary conditions to the solutions.
For the case of \eqn{Atrans} the scalar $\Phi$ denotes any component
of $A^\bot_\m$.
In order to cast \eqn{Azeom} into the form \eqn{fiiin}, we have
used \eqn{hgde} and defined $\Phi=e^C A_z$.
For $\b_i=\b_j$ we recover from \eqn{broken} eq. \eqn{unbroken} that
describes the cases with unbroken symmetry.
Hence, for full generality, we may use \eqn{broken}
in order to calculate current-current correlators.
We will follow the standard procedure of \cite{gkp,witten}
and we will work in Euclidean signature unless stated otherwise.

In order to proceed we need a complete set of eigenfunctions of
\eqn{broken}, which for the examples
we will discuss in the next section can be found explicitly and
is given in terms of hypergeometric functions.
Furthermore, we keep the solutions that blow up at the AdS
boundary since they correspond to current operator
insertions \cite{gkp,witten}.
Finally, we have to evaluate
the on shell-value of the action ${1 \ov \kappa^2} \int d^5x {\cal L}$
with ${1 \ov \kappa^2} = \frac{N^2}{16 \pi^2}$ for solutions
$\Phi$ of \eqn{broken}.\footnote{The overall normalization is found by
carefully keeping track of all the prefactors in the dimensional reduction
in the $S^5$-directions of the ten-dimensional type-IIB action
to five dimensions. In particular, ${1\ov \kappa^2} = {V_{S^5}\ov
4 \kappa_{10}^2} R^{8}$. Then using
$2 \kappa_{10}^2 = (2\pi)^7 \a'^4 g_s^2$, $R^4 = 4\pi g_s \a'^2 N$ and
$V_{S^5}=\pi^3$ we find the result mentioned above.}
We find the boundary term
\be
- \lim_{\epsilon \rightarrow 0} \frac{N^2}{32 \pi^2} e^B \Phi \del_z \Phi
\Big |_{z=\epsilon}^{z_{\rm max}} \equiv \frac{N^2}{16 \pi^2} k^2 H(k)   \ .
\label{onshell}
\ee
In order to keep formulas short in later sections we have
written out the overall factor $1/\kappa^2$ in the definition of
$H(k)$.
In order to obtain the correct result we have to normalize
$\Phi |_{z=\epsilon} = 1$ and take the limit in \eqn{onshell}.
Re-introducing Lorentz and group theory indices properly,
we can present the current-current correlators in momentum space
schematically as
\be\label{momcorr}
\langle J_\m^a(k) J_\n^b(-k) \rangle = \frac{N^2}{8 \pi^2} \delta^{ab}
\left(\delta_{\m\n} - \frac{k_\m k_\n}{k^2}\right) k^4 \tilde{G}(k)\ ,
\ee
where a group theory factor and the momentum space
version of the projector, which guarantees that
the amplitude is transverse, have been included.
The factor $H(k) \equiv k^2 \tilde{G}(k)$ depends also on the
adjoint indices $a,b$, but for reasons similar to those explained in footnote
2 we have not explicitly displayed them.

In the explicit calculations performed later in section 3 we will not
use $H(k)$ directly, as defined in \eqn{onshell}, because
the correlator in $x$-space is too singular to be Fourier transformed
to momentum space.
However, by using
differential regularization one can make sense of such
expressions by writing singular functions as derivatives of
less singular ones and then defining the Fourier transform by
formal partial integrations \cite{diffreg}.
In our case we have to take the correlator to
be of the form $\sim \square \square G(x)$ which is just
$k^4 \tilde{G}(k)$ in momentum space.
Hence, the correlator in $x$-space becomes
\be
\langle J_\m^a(x) J_\n^b(0)\rangle = \frac{N^2}{32 \pi^4}
\d^{ab}(\square \delta_{\m\n} - \del_\m \del_\n) \square G(x)\ ,
\label{cor1}
\ee
where
\be
G(x) = {1\ov 4\pi^2} \int d^4 k e^{i k\cdot x} {H(k)\ov k^2} = {1\ov r}
\int^\infty_0 dk H(k) J_1(k r)\ ,
\label{cor2}
\ee
with $J_1(kr)$ being a Bessel function.

\subsection{Supersymmetric quantum mechanics}

In this subsection we want to study general aspects of
the fluctuation equation \eqn{broken}, before we proceed in section 3
to describe two special cases where calculations can
be performed exactly.
Writing $\Phi = e^{-B/2} \Psi$ the field equation \eqn{broken}
turns into the one-dimensional Schr\"odinger equation
\be
-\Psi^{\prime\prime} + V \Psi = m^2 \Psi\ ,
\ee
with potential
\be
V={1\ov 4} (B^\prime)^2 +\ha B^{\prime\prime}
+ g^2 e^{2(A+\b_i+\b_j)} \sinh^2(\b_i-\b_j)\ .
\label{kkk1}
\ee
This potential, though not at all obvious, can be cast into a
form that appears in
supersymmetric quantum mechanics.
First, we rewrite it differently using the properties of our
solution \eqn{pro1} and \eqn{proo} and in particular \eqn{hd3}
which proves useful in turning derivatives with respect to the variable
$z$ into functions of the auxiliary function $F$ only:
\be
V  = {g^4 f^{1/2}\ov 64} \left[8 \sum_{i=1}^{6} {1\ov (F-b_i)^2}
- \Big(\sum_{i=1}^{6} {1\ov F-b_i}\Big)^2 \right]\ .
\label{hf8s}
\ee
Comparing with eq. $(4.16)$ of \cite{bbs} (after setting in there the
parameter $\Delta=4$) we find that this can be written solely in terms of
the conformal factor in the metric ansatz \eqn{metriki}
\be
\label{smth}
V=  {9\ov 4} {A^\prime}^2 - {3\ov 2} A^{\prime\prime} \ .
\label{kkk2}
\ee
This potential\footnote{An
alternative way to prove the equivalence of the potentials
\eqn{kkk1} and \eqn{kkk2} is to use the differential equation obeyed
by the $\b_i$'s, namely $\b_i^\prime =A^\prime +{g\ov 2} e^{A+2\b_i}$
\cite{bakas1}.} has the same form as the potential appearing
in supersymmetric quantum mechanics \cite{SQMwit,SQMrev}
with superpotential $W=-3/2 A^\prime$.
In fact, it is the
supersymmetric partner of the potential
\be
V_s= {9\ov 4} {A^\prime}^2 + {3\ov 2} A^{\prime\prime} \ ,
\label{jdhf1}
\ee
that appeared in studies of 2-point
functions for scalar fields or transverse graviton fluctuations
\cite{fgpw2,brand2,bakas1,cglp,DeWolfe0,bbs}; the relation of \eqn{jdhf1} to
supersymmetric quantum mechanics in the context of gauged supergravity
was first hinted in \cite{bakas1} and explicitly noted in \cite{DeWolfe0}.
Note that, the Schr\"odinger problem is universal and
does not depend on the indices $i,j$ of the gauge currents.
Consequently, the mass spectrum is the same irrespectively of whether it is
associated to
currents corresponding to broken or unbroken symmetries.
Instead, the wavefunction
$\Phi$ does depend on the indices $i,j$
through the explicit dependence on them of the conformal factor $B$
defined in \eqn{h93} (cf. footnote 3).

It is well known from the general theory of supersymmetric quantum mechanics
that the spectra of superpartner potentials, such as \eqn{smth}
and \eqn{jdhf1},
are identical except for a zero mode. However, in our case
such a mode is not normalizable
due to the asymptotic behavior of the function $A(z)$ as $z\to 0$
and, therefore, is not included in the spectrum. Hence, the spectra of
current fluctuations, corresponding to \eqn{smth} and those for dilaton
and transverse graviton fluctuations, corresponding to
\eqn{jdhf1}, exactly coincide, as advertised in the introduction.
We note, that related observations concerning a $SO(3)$ invariant
sector of 5d gauged supergravity and a particular
Coulomb branch flow have been made in \cite{massimo}.
\footnote{The authors of \cite{massimo} informed us that their
arguments concerning graviphotons are actually broader and
include all massive cases where $U(1)_R$ is broken.}

The analysis of the qualitative features of the spectrum can be done in
a similar fashion as in the case of the superpartner potential
arising in the case of scalar correlators \cite{bakas1,cglp}. At the boundary
$z=0$ the potential goes to $+\infty$ as $V\simeq {3\ov 4 z^2}$.
The behavior in the interior depends on the number
$n$ of constants of integration $b_i$ that equal the maximum constant among
them, $b_1$. We follow closely the discussion of \cite{bakas1,bbs} to which
we refer for further details.
For $n=4,5$ the range of $z$ necessarily extends to $+\infty$,
i.e. $0\leq z < \infty$, corresponding to $F=b_1$.
We find that, for $n=5$, the potential goes
to zero as $z\to \infty$ and the spectrum is continuous.
For $n=4$ the potential
approaches a constant value, as $z\to \infty$, which is given by
$V_{{\rm min}}={g^4\ov 4} f_0^{1/2} $.
Therefore, although the spectrum is continuous, there is a mass gap
whose squared
value is given by the minimum of the potential.
For $n=5$ the potential behaves as
\be
n=5: \qq
V_5 \simeq  {15/4\ov z^2}\ ,
\qq {\rm as} \quad z\to \infty\ .
\label{hjf24}
\ee
For $n=1,2,3$ the potential goes to $+ \infty$
as $F\to b_1$
and therefore the spectrum must be discrete.
Therefore there should be a
maximum value for $z$, denoted by $z_{\rm max}$, that is
determined by solving
the algebraic equation $F(z_{\rm max}g^2)=b_1$.
We find the behaviour
\ba
n=1,2,3: \qq
V_n &\simeq&  {C_{n}\ov (z-z_{\rm max})^2}\ ,
\qq {\rm as} \quad z\to z^-_{\rm max}\ ,
\nonumber\\
C_{n}&=& {4\ov (4-n)^2} -{1\ov 4}\ .
\label{hjf23}
\ea
For more details on the full
structure of the potentials \eqn{smth} and \eqn{jdhf1}, which
generically can be written using elliptic functions, the reader is referred
to the original literature \cite{bakas1,bbs}. In the two special cases,
to which we turn now in section 3, all
computations and results can be written in terms of elementary functions.

\section{The 2-point function}

In the previous section we introduced all necessary ingredients for
the calculation of correlators of symmetry currents
and pointed out the relation between supersymmetric quantum mechanics
and the fluctuation equations.
In this section we want to use these results and apply them
to two specific backgrounds worked out in \cite{fgpw2,brand1,bakas1}.
These backgrounds correspond
to distributions of D3-branes on a disc or a three-sphere \cite{kraus,sfe1}
and they both break the bulk gauge symmetry down to $SO(2) \times SO(4)$.
The broken symmetries form the coset $\frac{SO(6)}{SO(2) \times SO(4)}$.
On the dual field theory side these backgrounds correspond to states on the
Coulomb branch of ${\cal N} = 4$ SYM theory with reduced ${\cal R}$-symmetry.
In the following we will calculate the correlators in momentum and
position spaces.

\bigskip
\subsection{Distribution of D3-branes on a three-sphere}

We begin our exactly solvable examples with the case of a model representing
D3-branes uniformly distributed on a three-sphere. The expressions for
the metric and the scalar fields have been given in \cite{fgpw2,bakas1}.
The five-dimensional metric \eqn{metriki} has the conformal factor
\be
e^{2 A} = {r_0^2\ov R^2} {\cos^{2/3} u\ov \sin^2 u} \ ,\qq
0\leq u \leq {\pi\ov 2}\ ,
\label{mett1}
\ee
where we have defined for notational purposes the dimensionless
variable $u=r_0 z/R^2$. The parameter $r_0$ actually plays the r\^ole of the
radius of the three-sphere.
The $AdS_5$ boundary corresponds to $u=0$, whereas at $u=\pi/2$ there is a
naked curvature singularity. This is however naturally interpreted,
from a string theoretical point of view, as
the location of the distribution of the D3-branes on the three-sphere.

The profiles of the scalar fields are
\be
\label{profiles}
 e^{2\b_1}=e^{2\b_2} = \cos^{-4/3}u\ , \qq
 e^{2\b_3}= \dots =e^{2\b_6} = \cos^{2/3} u\ .
\ee
From a ten-dimensional view point, these scalars deform the
five-sphere line element that appears in the D3-brane solution
in such a way that the subgroup $SO(2)\times SO(4)$ of
the isometry group $SO(6)$ is preserved.  
The Schr\"odinger potential \eqn{smth} is found to be
\be
V={r_0^2\ov R^4}\left( -1 + {3\ov \sin^2 2u} \right)\ .
\ee
It is not difficult to show that
a complete orthonormal set of solutions to the corresponding
Schr\"odinger equation is given by
\be
\Psi_n =  \sqrt{{2 n(n+1)\ov n+1} {r_0\ov R^2}}\
{\cos^{3/2} u\ov \sin^{1/2} u}\
P_n^{(-1,1)}(\cos 2 u)\ ,\qq 0\leq u\leq {\pi
\ov 2 }\ , \qq n=1,2,\dots\ ,
\label{sooll1}
\ee
where the $P^{(-1,1)}_n$'s are Jacobi polynomials,
provided that the spectrum is given by
\be
 m^2_n =  {4r_0^2\ov R^4} n(n+1) \ ,\qq n=1,2,\dots\ ,
\label{llso2}
\ee
Note that the case with $n=0$, giving rise to a zero-mass eigenvalue,
is not included in the spectrum since the corresponding Schr\"odinger
norm diverges. The eigenvalues \eqn{llso2} coincide
with those found
for dilaton fluctuations in \cite{fgpw2,brand1} using the same background
as here, in agreement
with our general discussion in section 2.
Also the $n$-dependent overall constant in \eqn{sooll1} has
been chosen such that the $\Psi_n$'s are normalized to one.

The conformal factor appearing in the equation of the
fluctuations \eqn{broken} is:
\be
e^B = {r_0\ov R} \times \left\{\begin{array}{ll}
{\cos^3 u \ov \sin u} \ , & i,j=1,2 \ ,
 \\
{1 \ov \sin u\cos u} \ , & i,j =3,4,5,6  \ ,
\\
{\cos u\ov \sin u}\ , & i=1,2\ ,\quad j=3,4,5,6 \ .
\end{array}
\right.
\label{limi16}
\ee

\subsubsection{The 2-point functions}

Using \eqn{broken}, \eqn{profiles} and \eqn{limi16} we find the
wave equation for the transverse modes of the gauge field
in the unbroken $SO(2)$ subgroup,
the coset and the unbroken $SO(4)$:
\ba
(1-x) (x^2 \Phi^\prime)^\prime - \frac{\tilde{k}^2}{4} \Phi & = & 0 \ ,
\nonumber \\
(1-x) (x \Phi^\prime)^\prime - \frac{\tilde{k}^2}{4} \Phi -
 \frac{1-x}{4 x} \Phi & = & 0\ ,\qq
x \equiv \cos^2 u \in [0,1] \ , \\
x (1-x) \Phi^{\prime \prime} - \frac{\tilde{k}^2}{4} \Phi & = & 0 \ ,
\nonumber
\ea
where the prime denotes derivatives with respect to $x$
and $\tilde{k}^2 = R^4/r_0^2 k_\m k^\m$, i.e. is the length-square
of the four-vector
$k^\m$ rescaled for notational convenience with the indicated factor.

The wave-functions that blow up at the boundary at $x=1$
and are regular at the singularity at $x=0$ are given in terms of a
hypergeometric function as\footnote{Throughout
the paper we will make use of special functions and their properties
following the conventions of \cite{tipologio}.}
\be
\Phi = \G\big( (3+\D)/2\big) \G\big( (3-\D)/2\big) x^\l
F\left( \frac{1+\D}{2}, \frac{1-\D}{2},2,x \right) ~.
\label{sooll}
\ee
where $\D = \sqrt{1-\tilde{k}^2}$, and where we have introduced the
parameter $\l= 0,\ha$ and $1$ for the currents corresponding to the unbroken
$SO(2)$, the broken coset and the unbroken $SO(4)$ symmetries, respectively.
The proportionality constant
in \eqn{sooll} has been fixed such that $\Phi(1)=1$ and hence
at the boundary the solution
becomes proportional to a $\d$-function,
i.e., fully localized operator insertion.
It is interesting to note that the wavefunctions $\Phi$ in all
three cases differ only by different powers of $x$. This is
related, as we have seen, to the fact that the mass spectra
for broken and unbroken currents are identical.
From \eqn{sooll} we extract
\ba
H(\tilde{k}) &= & \frac{1-\l}{\tilde{k}^2}
\ +\ 1/4 \Big(\psi\left((1+\D)/2\right) + \psi\left((1-\D)/2\right)+ 2 \g\Big)
\nonumber \\
& =& -{\l\ov \tilde{k}^2}
\ +\ {1\ov 2} \sum_{n=1}^\infty {2 n + \tilde{k}^2 \ov n
(4n(n+1) + \tilde{k}^2)}  \  ,
\label{add1}
\ea
which has a discrete spectrum of poles at $\tilde{k}^2 = - 4 n(n+1)$,
$n=1,2,\ldots$, corresponding precisely to the mass eigenvalues \eqn{llso2}.
However, if $\l \neq 0$, there is an additional pole at $\tilde{k}^2=0$.
We will comment on this in various places below.

The three correlators differ only in the coefficient of the
$1/\tilde{k}^2$ term.  In the case
of scalar correlators this would just give
a contact term and could be ignored, but in the case of the
symmetry-current correlators this has important consequences as we
will explain shortly.
Using \eqn{cor2} we obtain the following
exact expression for the function $G(x)$ in the correlator \eqn{cor1}:
\be
G(x)\ =\  \l \frac{r_0^2}{2 R^4}  \ln r
\ +\ {r_0 \ov 2 R^2 r} \sum_{n=1}^\infty {2 n+1\ov \sqrt{n(n+1)}} \
K_1\left(2\sqrt{n(n+1)}{ r r_0 \ov R^2}\right)\ ,
\label{ggg1}
\ee
where $K_1$ denotes the modified Bessel function and
in writing the term containing $\ln r$ we discarded an infinite
constant. We have also dropped a $1/r^2$ term, which, since
$\square 1/r^2 \sim \d^{(4)}(r)$,
contributes only contact terms to the correlator which we
consistently ignore.
Hence, we find
\be
\square G(x)\ =\ \l {r_0^2 \ov R^4 r^2}
 \ + \ {2 r_0^3 \ov R^6 r}
\sum_{n=1}^\infty (2n+1) \sqrt{n(n+1)}
K_1\left(2\sqrt{n(n+1)} {r r_0 \ov R^2}\right) \ .
\ee

Let us perform the consistency check that
for small $r$, or equivalently, in the limit $r_0\to 0$,
we should recover the conformal result. The dominant contribution
in this limit comes from the infinite sum which
can be approximated by an integral
\be
G(x) = {1\ov 2 r^2} \int_{1}^{1/r}  {dn\ov  n}+\dots
\simeq  -{1 \ov 4 r^2} \ln r^2 \ , \qq {\rm as}\quad  r\to 0 \ .
\label{cv1}
\ee
This gives rise to
\be
\label{bfsphere}
\square G(x) \simeq {1\ov r^4}\ ,\qq {\rm as}\quad  r\to 0\ ,
\ee
which in turn, gives a $1/r^6$ fall off for the
correlator \eqn{cor1} at short distances. As expected, this
coincides with the result in the conformal case (see, for instance, eq. (30)
of \cite{fmmr}).

The behavior of $G(x)$ for large $r$ is easily
found from the asymptotic expansion of the modified Bessel function.
For large $r$ each separate term in the infinite sum behaves as
$e^{-m_{n} r}/r^{3/2}$, where $m_n$ are the mass eigenvalues in \eqn{llso2}
and hence gives rise to an exponential fall off. Keeping the two most
dominant contributions in the right hand side of \eqn{ggg1} we obtain
\be
G(x)\ \simeq\
 \l  {r_0^2\ov 2 R^4} \ln r\ +\ {3 \sqrt{\pi}\ov 8 \sqrt{2}}
\left(R^2\ov r_0 r\right)^{3/2}
e^{-2 \sqrt{2} r_0 r/R^2}\ , \qq {\rm as}\quad  r\to \infty \ .
\label{jke}
\ee
For the cases corresponding to the broken coset currents and the unbroken
$SO(4)$ currents we have $\l\neq 0$ and therefore the dominant
contribution for large $r$ comes from the first term in \eqn{jke}.
When substituted into the correlator in \eqn{cor1} it produces
a contact term, which we drop, and a term of the form
\be
\label{goldy}
\langle J_\m^a(x) J_\n^b(0)\rangle \simeq \l \d^{ab}
\frac{N^2}{4\pi^2}   { r_0^2 \ov R^4}
\frac{ 1}{r^6} \left(r^2 \delta_{\m\n} - 4 x_\m x_\n \right) \ ,
\qq {\rm as} \quad r\to \infty\ .
\ee
This term decays only with the forth power of the distance and
at first sight it might
be tempting to interpret it as arising from the massless
Goldstone boson associated with the broken symmetry.\footnote{Work on the
AdS/CFT correspondence and the Goldstone bosons has been reported
using a different model in \cite{dz2}.} From a physical point of view
there are several problems with such an interpretation:
First, this term does not appear on equal footing for all three types
of currents although they reside in the same supersymmetry multiplet.
Its existence might seemingly be acceptable or even desirable for the
broken symmetry, but this term also appears for the unbroken
$SO(4)$-symmetry currents. We also know from section 2 that the gauge
fields dual to the broken currents become massive via the Higgs mechanism
and, therefore, are not expected to produce any massless states.
Second, the pole of the massless state corresponds
to a non-normalizable mode
and it is not expected to show up in the two-point function.
The most plausible solution seems to be that these poles are
actually unphysical and should be dropped from the correlators.
Note that a similar problem
was found in \cite{DeWolfe} for the two-point function of
active scalars in the same backgrounds we are discussing here.
The mysterious massless poles in that paper were later shown
to be absent if a different prescription for the correlators is
used \cite{AFT}. It seems likely, although we have not checked,
that an improved prescription would resolve the puzzle in our
case as well.\footnote{Actually, we were able to explain the presence
of these massless poles we found in the supergravity calculation by
a field theory calculation in the free field approximation. These
results are added as an addendum at the end of this paper, since they
were found after publication of the original version of the
paper.}

\subsection{ Distribution of D3-branes on a disc}

Our second exactly solvable model represents
D3-branes uniformly distributed on a disc of radius $r_0$. The expressions for
the metric and the scalar fields have been given in \cite{fgpw2,bakas1}.
The five-dimensional metric \eqn{metriki} has the conformal factor
\be
e^{2 A} = {r_0^2\ov R^2} {\cosh^{2/3}u\ov \sinh^2 u} \ , \qq
0\leq u < \infty\ ,
\label{mett2}
\ee
where as before $u= r_0 z/R^2$.
The scalar fields are given by
\be
e^{2\b_1}= \dots =e^{2\b_4} = \cosh^{2/3} u\ ,
\qq e^{2\b_5}=e^{2\b_6} = \cosh^{-4/3} u\ .
\ee
As before, from a ten-dimensional type-IIB view point,
these scalars deform the
five-sphere line element that appears in the D3-brane solution
in such a way that the subgroup $SO(2)\times SO(4)$ of
the isometry group $SO(6)$ is preserved.  
The Schr\"odinger potential \eqn{smth} becomes
\be
V={r_0^2\ov R^4} \left( 1 + {3\ov \sinh^2 2 u} \right)\ .
\ee
The energy spectrum for this potential is continuous and has
a mass gap
\be
m^2  \geq  {r_0^2\ov R^4}\ .
\label{mmmg}
\ee
As before the zero mode corresponds to a non-normalizable wavefunction.

The conformal factor appearing in the equation of the fluctuations
\eqn{broken} is:
\be
e^B = {r_0\ov R} \times \left\{\begin{array}{ll}
{\cosh^3 u\ov \sinh u} \ , & i,j=1,2 \ , \\
{1\ov \sinh u\cosh u} \ , & i,j =3,4,5,6  \ , \\
{\cosh u\ov \sinh u}\ , & i=1,2 \ ,\quad j=3,4,5,6 \ .
\end{array}
\right.
\label{li16}
\ee

\subsubsection{The 2-point function}

The wave equation \eqn{fiiin} for the gauge fields of the unbroken $SO(2)$,
the broken coset and the unbroken $SO(4)$ symmetries, respectively, are:
\ba
x^2 (1-x) \Phi^{\prime\prime} -{\tilde k^2\ov 4} \Phi & = & 0 \ ,
\nonumber\\
x(1-x)(x \Phi^\prime)^\prime - \frac{\tilde k^2}{4} \Phi -
\frac{1}{4}(1-x) \Phi & = & 0 \ ,\qq
x \equiv {1 \ov \cosh^2 u} \in [0,1] \ ,
\label{gh9}\\
(1-x) (x^2 \Phi^\prime)^{\prime} - {\tilde k^2\ov 4} \Phi & = & 0\ ,
\ea
where, as before, $\tilde k^2 = k^2 R^4/r_0^2$.
The properly normalized solution that is also regular in the interior is
\be
\Phi =  {\G\big((1+\D)/2\big) \G\big( (3+\D)/2\big)\ov \G(1+\D)}
x^{(1+\D)/2-\l} F\left( \frac{\D-1}{2},\frac{\D+1}{2}, 1+\D,x \right)\ ,
\ee
where $\D = \sqrt{\tilde k^2+1}$ and similarly to before,
the parameter $\l= 0,\ha$
and $1$ for the currents corresponding to
$SO(2)$, to the coset and to $SO(4)$, respectively.
From this we obtain
\ba
H(\tilde k) & = & {\l-1 \ov k^2}\ + \ \ha \left( \psi\left((1+\D)/2\right)
+ \gamma\right)
\nonumber\\
& = & {\l-1 \ov k^2}\ + \
\ha \int_0^\infty dt {e^{-t}-e^{-{\D+1\ov 2} t}\ov 1-e^{-t}}\
\label{add2}
\ea
and then
\be
G(x)\ =\ \ha (1-\l) {r_0^2\ov R^4}\ln r \ + \
{1\ov 2 r^2} \int_0^\infty dy {y\ov \sinh y} {e^{-\sqrt{y^2 +
r_0^2 r^2/R^4}}\ov \sqrt{y^2 + r_0^2 r^2/R^4}}\ .
\ee
Using this result
it can be easily seen that the short distance behavior of the
propagator is the same as in the conformal case and in particular \eqn{cv1}
is recovered.
At large distances one finds that the two most dominant terms are
\be
G(x)\ \simeq \ \ha (1-\l) {r_0^2\ov R^4} \ln r \ + \ {\pi^2\ov 8} {R^2\ov
r_0 r^3} e^{-r_0 r/R^2}\ , \qq {\rm as} \quad r\to \infty\ ,
\ee
where naturally
the range of the Yukawa-term is set by the mass gap in \eqn{mmmg}.
Hence, for the case where $\l\neq 1$, corresponding to the cases of
the broken coset and the unbroken $SO(2)$ symmetries, the first term
dominates for large $r$ giving a contribution to the correlator
similar to \eqn{goldy}, but with $\l$ replaced by $1-\l$.
For similar reasons to those that we outlined
for the case of the sphere-distribution
of D3-branes after \eqn{goldy},
the interpretation of such a term as being related to
the Goldstone bosons is problematic and we believe that they
are unphysical. (However, see footnote 9.)

\section{Discussion}

In this letter we studied ${\cal R}$-symmetry current
correlators in certain states on the Coulomb branch of ${\cal N} = 4$
SYM using the standard description of the AdS/CFT correspondence.
The surprising result is that the spectra derived from the analytic
structure of the correlators agree with spectra of other operators
corresponding to dilaton and to the transverse graviton fluctuations.
Furthermore, it turned out that the spectra are identical and do not
depend on whether they are in the unbroken part of the left over
global symmetry or reside in the coset, except for certain zero-mass
poles which do depend on the sector. These poles give rise to
a $1/r^4$ fall off of the correlators at large distances, the
behavior expected of massless scalars, but we did not find good
physical reasons to identify them with Goldstone bosons of the
broken symmetry currents. We rather think that these poles are
unphysical since they correspond to non-normalizable states and
are inconsistent with the fact that the currents are all in the same
supersymmetry multiplet.\footnote{
After this paper was published in JHEP
we found convincing evidence that these poles are actually physical.
See footnote 9 on page 13 and especially the addendum at the end
of the paper for more details on the resolution of this puzzle.}

Rephrasing the fluctuation equations into a supersymmetric quantum mechanics
problem
we found that they all fall into the same universality class and, furthermore,
the Schr\"odinger potential are the supersymmetric partner potentials
arising from the dilaton or from the transverse graviton fluctuations, which
are identical. This indicates that all fluctuations
in such backgrounds fall into the same class of supersymmetric quantum
mechanics problems.

To obtain a more complete picture including the Goldstone bosons one
probably has to include additional modes that live on the D3-branes
which create the singularity in the infrared. In our set up with a continuous
distribution of branes
this seems a formidable task, and as a starting point it seems
more feasible to study simpler examples, e.g.
two stacks of coinciding branes or a single
test brane separated from a stack of branes, in which case one would
readily know the additional modes and their respective couplings to
the bulk fields. We leave these issues for future work.

It will also be interesting to investigate current-correlators using
solutions of $D=7$ and $D=4$ gauged supergravity that are dual to the (2,0)
theories in six dimensions and the three-dimensional theories with sixteen
supercharges, respectively,
on the Coulomb branch. For a class of such backgrounds
corresponding to a scalar-gravity sector analogous the one used in the present
paper the most general solution has been found and is very similar to that
in \eqn{metriki}-\eqn{hd3} \cite{bbs} (see also \cite{cglp}).
The spectrum, of fluctuations
corresponding to a massless scalar has been also exhaustively studied and
in some cases the computations can be performed explicitly \cite{bbs}.
Similarly to the present paper, in these cases as well,
it is quite plausible that the current-correlators
and the associated spectra are related
via supersymmetric quantum mechanics to those of the massless scalar.

\section*{Acknowledgements}

A.B. would like to thank Y.~Oz, J.~Sonnenschein
and S.~Yankielowicz for discussions,
and the universities of Tel-Aviv, Neuch\^atel
and Ludwig-Maximilians in Munich
for hospitality and financial support.
K.S. would like to thank the Theory Division at CERN for hospitality and
financial support during a considerable part of this research.
The research of K.S. was supported
by the European Union under contracts
TMR-ERBFMRX-CT96-0045 and -0090, by the Swiss Office for Education and
Science, by the Swiss National Foundation and by the contract
HPRN-CT-2000-00122.



\section*{Addendum}

The purpose of this addendum is to investigate the structure of the massless
poles that appear in 2-point functions of
broken symmetry currents in ${\cal N}=4$ SYM theory using purely field
theoretical techniques and to compare the results with those obtained
in section 3 using
supergravity and the AdS/CFT correspondence.
We had completed the essential part of this work around March of 2001.
Parts of it are based on ideas
developed around that time in collaboration with D. Freedman and K.
Skenderis.

\subsection*{General formulation }

We start with the case of unbroken $\cal R$-symmetry where the vev's
corresponding to the six scalars of the theory are turned off.
The $\cal R$-symmetry currents
$J^a_\m$ are represented as bilinears in the scalar fields
$X^i$, $i=1,2,\dots , 6$ transforming in the adjoint of $SU(N)$
\be
J^a_\m = {1\ov g_{\rm YM}^2} T^a_{ij} {\rm Tr}( X^i\del_\m X^j)\ + \
{\rm fermions}\ ,
\label{hl1}
\ee
where $T^a$ are $6\times 6$ matrices of $SO(6)$.
The scalars $X^i$, being free fields, obey the following two-point function
\footnote{In our conventions the field theory action has an overall
factor of $1/g_{YM}^2$.}
\be
\langle X^i_{pq}(x) X^j_{rs}(0)\rangle = g_{\rm
YM}^2 \d^{ij} (\d_{qr} \d_{ps} -
{1\ov N} \d_{pq} \d_{rs}){1\ov r^2}\ , \qq p,q,r,s=1,2,\dots , N\ .
\label{twopoint}
\ee
After performing the Wick contractions we compute the two-point
function for the currents
\be
\langle J_\m^a(x) J_\n^b(0)\rangle \sim N^2
\d^{ab}(\square \delta_{\m\n} - \del_\m \del_\n) {1\ov r^4}\ ,
\label{cor1ne}
\ee
where we have kept only the leading term in the
$1/N$-expansion.\footnote{For finite $N$, the $1/N$-term in \eqn{twopoint}
induces a shift which replaces the coefficient $N^2$ by $N^2\!-\!1$
corresponding to the dimension of the $SU(N)$ group. We also note that
the contribution of the fermions only affects the result by an
overall $N$-independent numerical constant
which is not important for our purposes.}
This is indeed the correct result for the two point function which
also agrees with the AdS/CFT result \cite{fmmr}.

In the case that the symmetry is broken by turning on non-zero scalar
vev's, we replace $X^i $ by $X^i_{vev} + \delta X^i$, where the $\delta X^i$
have the same free field two-point function as in \eqn{twopoint}.
Besides the bilinear
term \eqn{hl1} the current contains now a term linear in fluctuating
fields
\be
\delta J^a_\m = {1\ov g_{\rm YM}^2}
T^a_{ij} {\rm Tr}( X_{\rm vev}^i\del_\m \delta X^j)\ ,
\label{hl11}
\ee
where we have introduced the vev's
\be
X^i_{\rm vev}=\langle X^i\rangle = {\rm diag}(X^i_1,X^i_2,\dots , X^i_N)\ , \qq
\sum_{p=1}^N X^i_p=0\ .
\label{vvee}
\ee
At this point it is convenient to replace the adjoint $SO(6)$ indices by
$a=[ij]$ and $b=[kl]$. Then, the matrix elements of the
$SO(6)$ generators become
$T^{ij}_{mn}= \d_{im} \d_{jn} -\d_{jm} \d_{in}$.
The leading order correction to the conformal result \eqn{cor1ne} is
\be
\langle \delta J_\m^{ij} (x) \delta J_\n^{kl}(0)\rangle  \sim  {1\ov
g^2_{\rm YM}}
H^{ij,kl} \del_\m \del_\n {1\ov r^2}\ ,
\label{corrector}
\ee
where the group theoretical factor $H_{ij,kl}$ takes the form
\be
H^{ij,kl}  =  \d_{ik} A_{jl} - \d_{jk} A_{il} - \d_{il} A_{jk} +
\d_{jl} A_{ik} \ ,\qq  A_{ij} = \sum_{p=1}^N X^i_p X^j_p\ .
\label{ggrop}
\ee
It is clear that, in the UV where the vev's can be neglected, the conformal
result \eqn{cor1ne} dominates, whereas in the IR the dominant term is
\eqn{corrector}.
The symmetric tensor $A_{ij}$ is given in terms of the
scalar vevs only and depends on their distribution. In the
following we think of the vevs $X^i_{vev}$ as defining $N$ points in
${\bf R}^6$.
In most examples we use the fact that in the large $N$ limit such a
discrete distribution can be well approximated by a
continuous one.\footnote{This is correct as long as we work with energies
(distances in the gravity side) $U$ not too close to the vev values.
Typically the condition to be fulfilled for the continous approximation to
be valied is $U/X_{\rm vev}-1\gg {\cal O}(1/N)$, where $X_{\rm vev}$ is
a typical scalar vev value \cite{sfe1}.}
Furthermore, we will consider situations where the distribution spans only
a lower dimensional submanifold embedded in ${\bf R}^6$.
The tensor $H^{ij,kl}$ contains all the important information about
the zero mass poles. It is antisymmetric in the indices $ij$ and $kl$
separately and symmetric under pairwise exchange.
Note that only
if both indices $i,j$ are along the vev-distribution $A_{ij}$
is non-zero. That implies that if all indices correspond to directions
which are perpendicular to the distribution, then $H^{ij,kl}=0$.

\subsubsection*{Basic examples}

We digress to present a toy example of a discrete distribution of vevs in an
$N$-polygon enclosed by a ring of radius $r_0$ in the $1$-$2$ plane \cite{sfe1}
\be
X^i_{vev} = \left( r_0 \cos \phi_p, r_0 \sin \phi_p,0,0,0,0 \right) \ ,
\quad \phi_p=2\pi p/N\ ,\quad p=1,2,\dots , N\ .
\label{lv}
\ee
Computing the matrix elements $A_{ij}$ using the definition \eqn{ggrop}
is straightforward. We find that the
only non-zero components are $A_{11} = A_{22} = N r_0^2/2$.\footnote{
We have used the fact that
\be
\sum_{p=1}^N \cos^2(2 \pi p/N) = \sum_{p=1}^N \sin^2(2\pi p/N) =  N/2\ ,
\qq \sum_{p=1}^N \cos(2\pi p/N) \sin(2\pi p/N) =0\ .
\ee
}
We note that in this case we obtain the same result even if we approximate the
discrete distibution by a continous uniform distribution of vev's on the
circumference of the circle.

We now turn to the specific examples
of the distribution of vev's on a disc and on a three-sphere, which
we considered in section 3 using AdS/CFT correspondence.
In these cases a direct comparison with the free field calculation
can be performed.
In particular, in accordance with the convention in \eqn{momcorr}-\eqn{cor2},
the momentum space version of \eqn{corrector} can be
expressed in terms of a function $H(k)$
\be
H(k) \sim - \frac{1}{g^2_{\rm YM} N^2} \frac{H^{ij,kl}}{k^2} \ .
\label{hhh}
\ee

For the distribution on a three sphere it is obvious that
$A_{ii}=N r_0^2/4$, for $i=1,2,3,4$ and zero otherwise. These
results are most easily derived
in the continous approximation of the distributions.
Hence, using \eqn{hhh} and the facts that $g_{\rm YM}^2=g_s$ and
$R^4=4\pi g_s N$, we obtain
\be
H(k) \sim - \frac{r_0^2}{R^4} \frac{\lambda}{k^2}\ ,
\ee
where the parameter $\lambda = 0, \frac{1}{2}$ and $1$ corresponds to
currents in the transverse direction (unbroken $SO(2)$), broken currents
in the coset and directions along the distribution (unbroken $SO(4)$),
respectively. This agrees nicely with the AdS/CFT result \eqn{add1}.

For the distribution on a disc we have similarly that
$A_{ii}=N r_0^2/4$, for $i=1,2$ and zero otherwise.
Using \eqn{hhh} we compute
\be
H(k) \sim \frac{r_0^2}{R^4} \frac{\lambda - 1}{k^2}\ ,
\ee
where the parameter $\lambda = 0, \frac{1}{2}$ and $1$ corresponds to
currents along the distribution (unbroken $SO(2)$), broken currents
in the coset and directions orthogonal to the distribution (unbroken $SO(4)$),
respectively. Again we find precise agreement with the AdS/CFT result
\eqn{add2}.

\subsection*{Generalization to a class of models}

A natural question is whether the agreement between field theoretical results
and those obtained from supergravity goes beyond the two specific examples we
considered in detail.
In fact, we may systematize our approach and show that
such an agreement persists for all
models with vev distributions corresponding to the five-dimensional
supergravity solution \eqn{metriki}-\eqn{hd3}.

On the supergravity side
the distribution of vev's is encoded in the harmonic function appearing in
the ten-dimensional metric describing the gravitational field of D3-branes.
In our cases the harmonic function is \cite{bakas1}
\be
H^{-1}  = {4 \ov R^4}  f^{1/2} \sum_{i=1}^6 {y_i^2\ov (F-b_i)^2}\ ,
\label{dhj1}
\ee
where $F$ is determined in terms of the six transverse coordinates
$y_i$ as a solution of the algebraic equation
\be
\sum_{i=1}^6  {y_i^2\ov F-b_i} =4 \ .
\label{jk4}
\ee
The harmonic function is in general
\be
H=\sum_{p=1}^N{4\pi g_s \ov |\vec y - \vec X_p|^4}\ ,
\ee
where the vev values $\vec X_p$ in \eqn{vvee} became
the centers of the harmonic function. In
the continous approximation this takes the form
\be
H=4\pi g_s\int d^6x {\r(x)\ov |\vec y-\vec x|^4}\ ,
\ee
where the density $\r(x)$ is normalized as $\int d^6 x \r(x)=N$.
We would like to compute $A_{ij}$ in \eqn{ggrop},
which for a continuous distribution reads
\be
A_{ij}= \int d^6x \r(x) x_i x_j\ .
\ee
In general this can be found from the large $r$ expansion
\be
H={R^4\ov r^4} - 4\pi g_s {2 \ov r^6}
\left( \d_{ij} -{6 y^i y^j\ov r^2}\right) A_{ij}
+\dots\ .
\ee

Returning to our cases where the harmonic function has the specific form
\eqn{dhj1}, we may cast its large $r$ expansion into the above form with
\be
A_{ij} = N b_{1j}\ \d_{ij}\ ,
\label{jefd}
\ee
where we define in general $b_{ij}=b_i - b_j$.
We see that our general distributions allow a diagonal matrix $A_{ij}$.
Hence, the only non-zero independent components of the group theoretical factor
$H^{ij,kl}$ are $H^{ij,ij}$.
If all indices correspond to directions which are perpendicular to the
distribution then $H^{ij,kl}=0$, whereas if
all directions are along the distribution
$H^{ij,ij}=N (b_{1j} + b_{1i})$.
If we are in the coset one index is along the distribution (say $i$) and one
is orthogonal to it (say $j$), then one of the above terms is missing
and therefore $H^{ij,ij}=N b_{1i}$. This agrees perfectly with the
two special
cases of the disc and sphere distribution that we considered before.

\subsubsection*{Correlators from supergravity}

Let us consider the equation \eqn{broken}
but in terms of the variable $F$
\be
{d\ov dF} \left( (F-b_i)(F-b_j)  {d\Phi \ov dF}\right) -
k^2 { (F-b_i)(F-b_j)\ov f^{1/2}}\Phi - {b_{ij}^2\ov 4 (F-b_i)(F-b_j)}\Phi
=0 \ ,
\label{hjg}
\ee
where $F$ was defined in equation \eqn{pro1}.
Equation \eqn{hjg} was solved exactly
for the cases of the disc sphere and the sphere distribution.
For the purposes of this
addendum it suffices to concentrate on the limit $k^2\to 0$,
where \eqn{hjg} can be solved exactly for any distribution.
This will give the leading contribution to the two-point function of
currents for large distances.
At the AdS boundary $F\to \infty$ we
impose the usual boundary condition
$\Phi\to 1$ corresponding to a point-like source.
Furthermore, we require $\Phi$ to be
smooth at the singularity $F=b_1$ in the interior. In the following we
use units where $g=2/R=1$.

\medskip
\no
\underline{Currents transverse to the distribution}:

In this case the indices of the current $i,j$  are such that
$b_i=b_j=b_1$. 
Demanding regularity at the singularity $F=b_1$
and imposing the normalization condition at the boundary gives
\be
\Phi=1\ .
\ee
Therefore \eqn{onshell} gives
\be H(k)=0\ .
\ee
As expected this agrees with the field theoretical result.

\no
\underline{Currents longitudinal to the distribution}:

In this case the indices of the current are such that $b_i, b_j\neq b_1$.
As before, regularity at the singularity at $F=b_1$
and the normalization condition at the boundary give
\be
\Phi = {1\ov b_{ij}} \left( b_{1j} \left({F-b_i\ov F-b_j}\right)^{1/2}
-b_{1i} \left({F-b_j\ov F-b_i}\right)^{1/2} \right)\ ,
\ee
from which we compute using \eqn{onshell} that
\be
H(k) = - {b_{1i}+ b_{1j}\ov 4 k^2}\ .
\ee
This is in perfect agreement with field theory expectations as spelled out
after \eqn{jefd}.
A particularly interesting case is when $b_i=b_j\neq b_1$. Then the
above expressions reduce to
\be
\Phi = {F-b_1\ov F - b_i}\
\ee
and
\be
H(k) = - {b_{1i}\ov 2 k^2}\ .
\ee
The case of the sphere and disc distributions correspond precisely to that
with $b_{1i}=r_0^2/4$ ($b_1$ can be put to zero by a shift of the coordinate
$F$), for $i=1,2,3,4$ and $i=1,2$, respectively.

\medskip
\no
\underline{Currents in the coset}:

In this case the currents indices are such that $b_i=b_1$ and $b_j\neq b_1$.
Proceeding as before we find that
\be
\Phi = \left({F-b_1\ov F-b_j}\right)^{1/2}
\ee
and that
\be
H(k) = - {b_{1i}\ov 4 k^2}\ .
\ee
Again, one easily sees that this agrees with field theoretical expectations.

\subsection*{Comments on the masses of gauge bosons}

Finally, we mention some usefull facts about the masses of the
W-bosons that arise on a generic point of the Coulomb branch
of the ${\cal N}=4$ SYM theory.
The general mass matrix is read off from eq. (63) of \cite{sfe2}
\be
(M^2)_{pq}=|\vec X_p-\vec X_q |^2,\qq p,q=1,2,
\dots , N\ ,
\ee
up to a numerical constant of order 1. Hence, the masses have the geometrical
interpretation as the
distances between the various vev positions distributed in the ${\bf R}^6$
scalar space.
Equivalently, they are given by
the masses of the strings stretched between the D3-branes located
at these points. Since we may shift uniformly all vectors
$\vec X_p$'s without changing the Physics, the number of elements are
in a generic case $N^2-1$ as it should be.
It is clear that depending on the specific vev distributions
some of these masses might be degenerate.
In particular, in the case of the discrete distribution of vev's in the
$N$-polygon we find, using \eqn{lv}, that (see eq. (66) of \cite{sfe2})
\be
M_n = r_0 \sin(\pi n/N)\ ,\qq n=1,2,\dots , N\ ,
\label{hwe}
\ee
which is an exact result for any $N$.
The degeneracy for the zero mode is $d_N=N-1$ and for the rest $d_n=2(N-n)$.
It is easily seen that $\sum_{n=1}^N d_n=N^2-1$.
Hence, for large $N$ there are W bosons with masses of order $r_0$
and light masses of order $r_0/N$.
In the case of vev's distributed on a disc a similar result can also be derived
starting from a discrete distribution  \cite{sfe1,brand1}
whose limit is the continuous one we have been using.


\begin{thebibliography}{99}

\bibitem{malda}
J.~Maldacena,
Adv. Theor. Math. Phys. {\bf 2} (1998) 231, {\tt
hep-th/9711200. }

\bibitem{gkp}
S.~S.~Gubser, I.~R.~Klebanov and A.~M.~Polyakov,
Phys.\ Lett.\  {\bf B428} (1998) 105,
{\tt hep-th/9802109}.

\bibitem{witten}
E.~Witten,
Adv.\ Theor.\ Math.\ Phys.\  {\bf 2} (1998) 253,
{\tt hep-th/9802150};
Adv.\ Theor.\ Math.\ Phys.\  {\bf 2} (1998) 505,
{\tt hep-th/9803131}.

\bibitem{MW}
J.A. Minahan and N.P. Warner,
JHEP {\bf 06} (1998) 005, {\tt hep-th/9805104}.

\bibitem{kraus}
P.~Kraus, F.~Larsen and S.~P.~Trivedi,
JHEP {\bf 9903} (1999) 003,
{\tt hep-th/9811120}.

\bibitem{sfe1}
K. Sfetsos,
JHEP {\bf 01} (1999) 015, {\tt hep-th/9811167}.

\bibitem{kw}
I.~R.~Klebanov and E.~Witten,
Nucl.\ Phys.\  {\bf B556} (1999) 89,
{\tt hep-th/9905104}.

\bibitem{KS1}
A. Kehagias and K. Sfetsos,
Phys. Lett. {\bf B454} (1999) 270, {\tt hep-th/9902125}
and
Phys. Lett. {\bf B456} (1999) 22, {\tt hep-th/9903109};
S.S. Gubser, {\it Dilaton-driven confinement},
{\tt hep-th/9902155};
H. Liu and A.A. Tseytlin,
Nucl. Phys. {\bf B553} (1999) 231,
{\tt hep-th/9903091};
N.R. Constable and R.C. Myers,
JHEP {\bf 11} (1999) 020,
{\tt hep-th/9905081}.

\bibitem{fgpw2}
D.~Z.~Freedman, S.~S.~Gubser, K.~Pilch and N.~P.~Warner,
JHEP {\bf 0007} (2000) 038,
{\tt hep-th/9906194}.

\bibitem{brand1}
A.~Brandhuber and K.~Sfetsos,
Adv. Theor. Math. Phys. {\bf 3} (1999) 851, {\tt hep-th/9906201}.

\bibitem{bakas1}
I.~Bakas and K.~Sfetsos,
Nucl.\ Phys.\  {\bf B573} (2000) 768,
{\tt hep-th/9909041}.

\bibitem{cglp}
M.~Cvetic, S.~S.~Gubser, H.~Lu and C.~N.~Pope,
Phys. Rev. {\bf D62} (2000) 086003, {\tt hep-th/9909121}.

\bibitem{bbs}
I.~Bakas, A.~Brandhuber and K.~Sfetsos,
Adv. Theor. Math. Phys. {\bf 3} (1999) 1657,
{\tt hep-th/9912132}.

\bibitem{gppz1}
L.~Girardello, M.~Petrini, M.~Porrati and A.~Zaffaroni,
JHEP {\bf 9812} (1998) 022,
{\tt hep-th/9810126}.

\bibitem{dz1}
J.~Distler and F.~Zamora,
Adv.\ Theor.\ Math.\ Phys.\  {\bf 2} (1999) 1405,
{\tt hep-th/9810206}.

\bibitem{khavaev}
A.~Khavaev, K.~Pilch and N.~P.~Warner,
Phys.\ Lett.\  {\bf B487} (2000) 14,
{\tt hep-th/9812035}.

\bibitem{KLM}
A. Karch, D. Lust and A. Miemiec,
Phys. Lett. {\bf B454} (1999) 265,
{\tt hep-th/9901041}.

\bibitem{fgpw1}
D.~Z.~Freedman, S.~S.~Gubser, K.~Pilch and N.~P.~Warner,
{\it Renormalization group flows from holography supersymmetry and a
c-theorem},
{\tt hep-th/9904017}.

\bibitem{Behrndt}
K. Behrndt, Nucl.\ Phys.\  {\bf B573} (2000) 127,
{\tt hep-th/9907070}.

\bibitem{gppz2}
L.~Girardello, M.~Petrini, M.~Porrati and A.~Zaffaroni,
Nucl.\ Phys.\  {\bf B569} (2000) 451,
{\tt hep-th/9909047}.

\bibitem{dz2}
J.~Distler and F.~Zamora,
JHEP {\bf 0005} (2000) 005,
{\tt hep-th/9911040}.

\bibitem{Polchinski:2000uf}
J.~Polchinski and M.~J.~Strassler,
{\it The string dual of a confining four-dimensional gauge theory},
{\tt hep-th/0003136}.

\bibitem{Pilch:2000ue}
K.~Pilch and N.~P.~Warner,
{\it N = 2 supersymmetric RG flows and the IIB dilaton},
{\tt hep-th/0004063}.

\bibitem{Brandhuber:2000ct}
A.~Brandhuber and K.~Sfetsos,
Phys. Lett. {\bf B488} (2000) 373, {\tt hep-th/0004148}.

\bibitem{Evans:2000ap}
N.~Evans and M.~Petrini,
{\it AdS RG-flow and the super-Yang-Mills cascade},
{\tt hep-th/0006048}.

\bibitem{Pilch:2000fu}
K.~Pilch and N.~P.~Warner,
{\it N = 1 supersymmetric renormalization group
flows from IIB supergravity},
{\tt hep-th/0006066}.

\bibitem{Klebanov:2000hb}
I.~R.~Klebanov and M.~J.~Strassler,
{\it Supergravity and a confining gauge theory:
Duality cascades and  (chi)SB-resolution of naked singularities},
JHEP {\bf 0008} (2000) 052,
{\tt hep-th/0007191}.

\bibitem{Buchel:2000cn}
A.~Buchel, A.~W.~Peet and J.~Polchinski,
{\it Gauge dual and noncommutative extension of an
N = 2 supergravity  solution},
{\tt hep-th/0008076}.

\bibitem{Evans}
N.~Evans, C.~V.~Johnson and M.~Petrini,
{\it The enhancon and N = 2 gauge theory/gravity RG flows},
{\tt hep-th/0008081}.

\bibitem{PPN}
M. Pernici, K. Pilch and P. van Nieuwenhuizen, Nucl. Phys. {\bf B259}
(1985) 460.

\bibitem{GRW}
M. Gunaydin, L.J. Romans and N.P. Warner, Phys. Lett. {\bf 154B}
(1985) 268 \break
and Nucl. Phys. {\bf B272} (1986) 598.

\bibitem{fmmr}
D.~Z.~Freedman, S.~D.~Mathur, A.~Matusis and L.~Rastelli,
Nucl.\ Phys.\  {\bf B546} (1999) 96,
{\tt hep-th/9804058}.

\bibitem{dhoker}
E.~D'Hoker and D.~Z.~Freedman,
Nucl.\ Phys.\  {\bf B544} (1999) 612,
{\tt hep-th/9809179}.

\bibitem{Anselmi}
D.~Anselmi, L.~Girardello, M.~Porrati and A.~Zaffaroni,
Phys.\ Lett.\  {\bf B481} (2000) 346,
{\tt hep-th/0002066}.

\bibitem{brand2}
A.~Brandhuber and K.~Sfetsos,
JHEP {\bf 9910} (1999) 013,
{\tt hep-th/9908116}.

\bibitem{DeWolfe}
O.~DeWolfe and D.~Z.~Freedman,
{\it Notes on fluctuations and correlation functions in
holographic  renormalization group flows}, {\tt hep-th/0002226};

\bibitem{AFT}
G.~Arutyunov, S.~Frolov and S.~Theisen,
Phys.\ Lett.\  {\bf B484} (2000) 295,
{\tt hep-th/0003116}.

\bibitem{massimo}
M.~Bianchi, O.~DeWolfe, D.~Z.~Freedman and K.~Pilch,
{\it Anatomy of two holographic renormalization group flows},
{\tt hep-th/0009156}.

\bibitem{OthersCoulomb}
M.~S.~Costa,
Phys. Lett.  {\bf B482} (2000) 287, {\tt hep-th/0003289}.
and M.~S.~Costa,
JHEP {\bf 0005} (2000) 041, {\tt hep-th/9912073};
R.~C.~Rashkov and K.~S.~Viswanathan,
Phys. Rev. {\bf D62} (2000) 046009, {\tt hep-th/9911160}.

\bibitem{cvetic}
M.~Cvetic, H.~Lu, C.~N.~Pope, A.~Sadrzadeh and T.~A.~Tran,
{\it Consistent SO(6) reduction of type IIB supergravity on S**5},
{\tt hep-th/0003103}.

\bibitem{diffreg}
D.~Z.~Freedman, K.~Johnson and J.~I.~Latorre,
Nucl.\ Phys.\  {\bf B371} (1992) 353.

\bibitem{SQMwit}
E. Witten, Nucl. Phys. {\bf B188} (1981) 513.

\bibitem{SQMrev}
For a review see, F. Cooper, A. Khare and U. Sukhatme,
Phys. Rept. {\bf 251} (1995) 26,
{\tt hep-th/9405029}.

\bibitem{DeWolfe0}
O.~DeWolfe, D.~Z.~Freedman, S.~S.~Gubser and A.~Karch,
Phys. Rev. {\bf D62} (2000) 046008, {\tt hep-th/9909134}.

\bibitem{tipologio}
I.S.~Gradshteyn and I.M.~Ryzhik,
{\it Table of integrals, series and products}, Fifth edition
(Academic Press, New York, 1994).

\bibitem{sfe2}
K.~Sfetsos,
{\it Dynamical emergence of extra dimensions and warped geometries},
Nucl. Phys. {\bf B} (in print), {\tt hep-th/0106126}.

\end{thebibliography}
\end{document}